\documentstyle[twoside,espcrc2-mod,epsfig]{article}

\newcommand{\B}{{\em BeppoSAX\/}}
\newcommand{\ie}{{\em i.e.,\/}\ }
\newcommand{\eg}{{\em e.g.,\/}\ }
\newcommand{\cgs}{\mbox{erg~cm$^{-2}$~s$^{-1}$}}
\newcommand{\pcs}{\mbox{ph~cm$^{-2}$~s$^{-1}$}}
\title{High energy properties of X--ray sources observed with \B}

\author{F. Frontera\address{%
  \ Dipartimento di Fisica, Universit\`a di Ferrara, Via Paradiso 12,
  44100 Ferrara, Italy}$^{\rm ,b}$,
D. Dal~Fiume\address{%
  \ Istituto Tecnologie e Studio Radiazioni Extraterrestri (TeSRE), C.N.R., \\
  \ \ via Gobetti 101, 40129 Bologna, Italy},
G. Malaguti$^{\rm b}$, L. Nicastro$^{\rm b}$, 
  M. Orlandini$^{\rm b}$, E. Palazzi$^{\rm b}$, E. Pian$^{\rm b}$,
F.~Favata\address{%
 \ Astrophysics Division, Space Science Department of ESA, ESTEC \\
 \ \ Keplerlaan 1, 2200 AG Noordwijk, The Netherlands},
A. Santangelo\address{%
  \ Istituto Fisica Cosmica e Applicazioni all'Informatica (IFCAI), C.N.R., \\
  \ \ via La Malfa 153, 90146 Palermo, Italy},
}
 
\begin{document}

\begin{abstract}
\noindent We report on highlight results on celestial sources observed in the
high energy band ($>20$~keV) with \B. In particular we review the spectral
properties of sources that belong to different classes of objects, \ie stellar
coronae (Algol), supernova remnants (Cas~A), low mass X--ray binaries (Cygnus
X--2 and the X--ray burster GS1826--238), black hole candidates (Cygnus X--1)
and Active Galactic Nuclei (Mkn~3). We detect, for the first time, the
broad-band spectrum of a stellar corona up to 100 keV;  for Cas~A we report
upper limits to the $^{44}$Ti line intensities that are  lower than those
available to date; for Cyg X--2 we report the evidence of a high energy
component; we report a clear detection of a broad Fe K line  feature from Cyg
X--1 in soft state and during its transition to hard state; Mkn~3 is one of
several Seyfert 2 galaxies detected with \B\ at high energies, for which
Compton scattering process is important.
\end{abstract}

\maketitle

\section{INTRODUCTION}

High energy properties of celestial X--ray sources give important information
to understand their radiation mechanisms and the energetic processes occurring
in them and/or in their environments.

Hard X--ray emission ($>20$~keV) is currently observed from several classes of
X--ray sources. Galactic X--ray sources that are known emitters of hard X--rays
include  black-hole candidates, X--ray pulsars, weak-magnetic-field neutron
stars in  Low Mass X--ray Binaries (LMXRBs), mainly X--ray bursters (XRBs), 
Cataclismic Variables (CV), in particular Polars, Crab-like supernova 
remnants.

High energy spectra of black-hole candidates (BHC) have permitted to 
infer the presence of Comptonization processes of soft photons
occurring close to the black-hole (\eg in a disk corona). 

XRBs, that are weakly magnetized neutron stars (with surface field intensity
$B\leq 10^{10}$--$10^{11}$~G), turned out to be hard X--ray emitters, once the
sensitivity of the high  energy instruments was increased at the 10~mCrab level
(see \cite{tavani97} for a recent review). From their spectral properties,
similarities with and diversities  from BHCs have been inferred, like the
presence of an accretion disk that can extend, as in the case of BHC, close to
the surface of the compact object, and the presence of an additional component
of soft photons that, unlike in BHCs, originates from the neutron star surface
and can be a major source of thermal emission and electron cooling through
Comptonization.

X--ray pulsars are well known emitters of hard X--rays. Observations in the
hard X--ray band are relevant in order to get a measurement of the magnetic
field intensity at the neutron star surface. Even if current models of the
X--ray spectrum of these objects are still unsatisfactory at high energies, the
measurement of cyclotron resonance features gives a direct estimate of the
intensity of the neutron star magnetic field strength~\cite{dalfiume97}.

Emission from young shell-like supernova remnants mainly extends to low
energies ($<20$~keV). Detection of hard X--rays with determination of their
spectral  properties can provide important information on the emission
mechanism  (thermal vs.\ non thermal, like synchrotron radiation).

%
%Figure 1
%
\begin{figure}
\centerline{\epsfig{figure=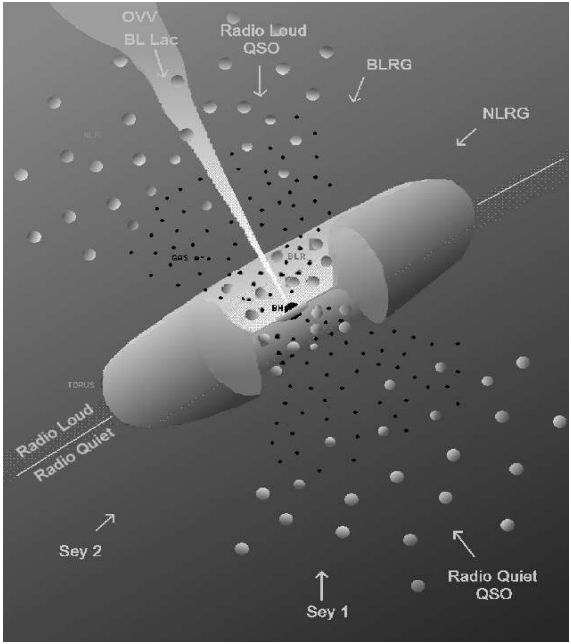,height=8.5cm,width=\columnwidth}}
\caption[]{Unified model for AGNs. Adapted from \cite{urry95}.}
\label{fig:fig_1}
\end{figure}

From stellar coronae, apart the Sun, hard X--rays have never been observed. As 
we will see, this gap has been filled with \B.

Among the extragalactic X--ray sources, hard X--ray emission is observed from
Active Galactic Nuclei (AGNs), that include Seyfert galaxies of both types (1
and 2), radio quiet QSOs and radio loud QSOs (which include blazars).  A great
effort is currently under way to interpret the different classes of AGNs in a
unified scheme, which is sketched in Fig.~\ref{fig:fig_1}. The basic energy
production mechanism is accretion of matter onto a massive black hole ($\approx
10^8$~M$_\odot$) via an accretion disk. A massive toroid of larger radius (in
the range from several parsecs to few tens of parsecs~\cite{krolik94}) is
assumed to surround the accretion disk. Depending on the configuration of the
disk with respect to the toroid, on their relative sizes and distances, and on
the viewing angle,   an AGN should show different observational features and
thus  fall in one of the different classes above mentioned.

The above scheme is being tested also for stellar-mass black holes  (\eg Cygnus
X--1). Thus the unified scheme can be a general picture to interpret galactic
and extragalactic black holes, accreting matter via an  accretion disk. Given
many similarities in the X--ray emission from stellar mass BHs  with
low-magnetic-field neutron stars in LMXRBs, the unified scheme now applied to
AGNs could be extended to several classes of X--ray sources. Hard X--ray
spectral properties of these sources can provide unique information to diagnose
the presence of a black hole versus a  weak-magnetic-field neutron star, to
test the unified model for AGNs and its validity for stellar mass BHCs.

Thank to a broad energy band of operation (0.1--300~keV) and a uniform flux
sensitivity in this range, \B\ \cite{BeppoSAX} has the unmatched capability of
simultaneously sampling the spectrum of X--ray sources over more than three
decades of energy.

The SAX/PDS instrument \cite{PDS}, with a sensitivity of about 1~mCrab at
100~keV, allows an accurate determination of the spectrum of many X--ray
sources at the highest energies (13--200~keV).

For the brightest sources ($>10$~mCrab), the HPGSPC instrument (6--60 keV) 
\cite{HPGSPC} provides the spectral coverage necessary to match the information
provided by the low energy instruments LECS (0.1--10~keV) \cite{LECS} and MECS
(2--10~keV) \cite{MECS} telescopes and the PDS.

Here we review some relevant results obtained with \B\ during its  Performance
Verification Phase and first year Core Program, with particular focus on the
PDS instrument. The spectral deconvolution was performed with the XSPEC
software package, by using the instrument response function distributed from
the \B\ Scientific Data Center.

\section{HIGHLIGHT RESULTS}

The reviewed sources span a large range of intensities and include a stellar
corona (Algol), a supernova remnant (Cas~A), two LMXRBs, one of which (Cygnus
X--2) is a Z source and the other (GS1826--634) is an X--ray burster, a BHC
(Cygnus X--1) observed in two different spectral states and the Seyfert 2
galaxy Mkn~3.

\subsection{Algol}

%
% Figure 2
%
\begin{figure}
\centerline{\epsfig{figure=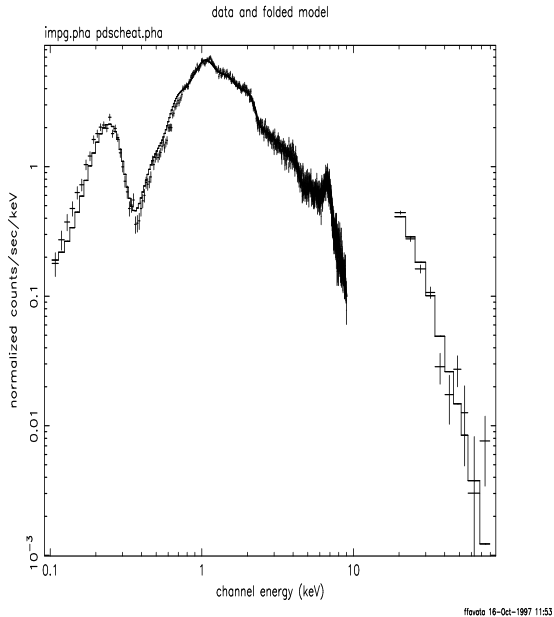,height=8.5cm,width=\columnwidth,angle=-90}}
\caption[]{Count rate spectrum of Algol during the maximum of a long flare as
observed by \B. Superposed is the best fit thermal model~\cite{mkl95}.}
\label{fig:fig_2}
\end{figure}

The general nature of stellar coronae as a class of thermal soft X--ray
emitters was already established with the {\em Einstein\/}
observatory~\cite{vaiana81}. The extensive observations with soft X--ray
telescopes ({\em Einstein\/}, ROSAT) has shown that the typical peak plasma
temperatures in the more active sources are of the order of a few keV, although
during intense flares their X--ray luminosity and coronal temperatures
increases strongly, leading to the expectation that hard X--rays may be
detected. This phenomenon has been confirmed during the recent observation with
\B\ of Algol~\cite{favata97a}. Algol is a binary system composed by a B8V and a
K2IV star, with a binary period of 2.9 days. \B\ observed the complete
evolution of a large flare, lasting about 1~day. During the flare the soft
(0.1--10 keV) X--ray luminosity increased by a factor more than 20. The source
spectrum measured with LECS and PDS instruments during the peak of the flare is
shown in Fig.~\ref{fig:fig_2}. The flare spectrum is reasonably described with
a two component emission model from a hot diffuse gas (MEKAL model in
XSPEC)~\cite{mkl95}. The preliminary analysis shows no evidence for the
presence of a non-thermal component, up to highest observed energies. The
characteristic temperatures of the flare spectrum are of the order of $\sim 2
\times 10^{7}$ and $\sim 1 \times 10^{8}~^\circ$K. Further details of this
observation are described in~\cite{favata97a}.

\subsection{Cas~A}

\begin{figure}
\centerline{\epsfig{figure=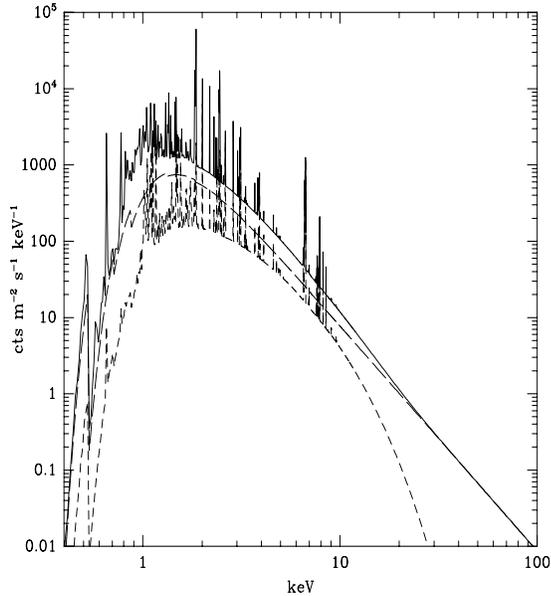,height=8.5cm,width=\columnwidth}}
\caption[]{Deconvolved spectrum of Cas A. The different components are shown.
From \cite{favata97b}.}
\label{fig:fig_3}
\end{figure}

The supernova remnant Cas~A was observed with \B\ on August 6, 1996. A
non-thermal high-energy component in the X--ray emission from the source has
been clearly detected~\cite{favata97b}. The broad-band (0.5--100~keV) X--ray
spectrum of the source (see Fig.~\ref{fig:fig_3}) is modeled using the sum of
three components: one Non-Equilibrium of Ionization (NEI) plasma component
representing the emission from the ejecta; one NEI component representing the
emission from the shocked material surrounding the circumstellar medium; a
power law (PL) component to model hard X--ray emission. Best fit parameter
values of the PL are a photon index of $2.95^{+0.10}_{-0.05}$ and a
normalization parameter of 0.69~ph~cm$^{-2}$~s$^{-1}$~keV$^{-1}$ at 1~keV. The
power law high energy component  is very likely of synchrotron
origin~\cite{allen97}. It gives a sizeable contribution at lower energies,
being comparable in intensity to the thermal continuum at the position of the
Fe K complex.

The search for radioactive emission lines due to $^{44}$Ti formed during the
supernova explosion gives a 2$\sigma$ upper limit to the intensity of the line
at 68~keV of $1.3 \times 10^{-5}$~\pcs\ and an upper limit a factor about 10
times lower ($4.4 \times 10^{-6}$~\pcs) at 78~keV. A 99\% upper limit of 8.6
$\times 10^{-5}$~\pcs\ for both lines was given by \cite{the95} with OSSE and a
similar value was obtained with RXTE~\cite{rothschild97}.

%
% Figure 3
%

\subsection{CYGNUS X--2}

The LMXRB Cygnus X--2 is a member of a binary system (binary period of 
9.84~days), consisting of a low-magnetic-field neutron star and a late type
low-mass ($\sim 0.7\,{\rm M_\odot}$) star V1341
Cyg~\cite{cowley79,mcclintock84,kahn84}. It is classified as Z source, on the
basis of its X--ray colour-colour spectral
behaviour~\cite{schulz89,wijnands97}. Its low-energy continuum spectrum,
measured with the EXOSAT satellite,  can be fitted with a Comptonization model
($kT = 3.7$~keV, optical depth $\tau = 9.4$) plus a black body component ($kT =
1.21$~keV)~\cite{white88}. An emission Fe K line was first detected with the
{\it Tenma}~\cite{hirano87} and EXOSAT~\cite{chiappetti90} satellites and,
later on, it was resolved with the Broad Band X--ray Telescope (BBXRT) which
detected a broad (${\rm FWHM} = 0.97$~keV) line with centroid energy 6.71 keV,
and ${\rm EW} = 60\pm 27$~eV~\cite{smale93}.

Past high energy observations of Cygnus X--2~\cite{matt90} showed that only a
small fraction (about 1\%) of the total source luminosity is emitted in hard
X--rays.

%
% Figure 4
%
\begin{figure}
\centerline{\epsfig{figure=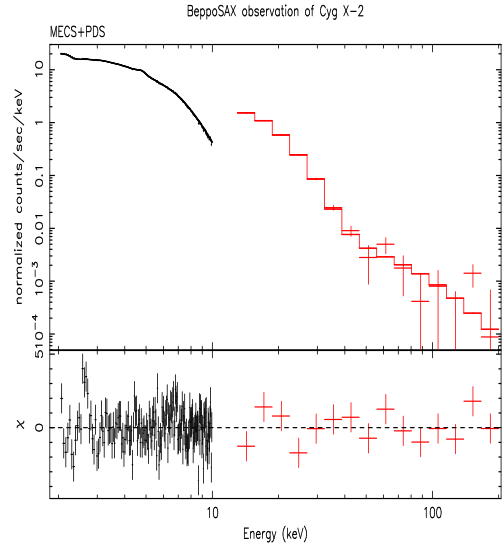,height=8.5cm,width=\columnwidth,angle=-90}}
\caption[]{Top: 2--200~keV count rate spectrum of Cygnus X--2 with superposed
the best fit model (see text). Bottom: residuals with respect to the model.}
\label{fig:fig_4}
\end{figure}

\B\ observed Cygnus X--2 on July 23, 1996, during the Science  Verification
Phase. Results obtained with the LECS instrument have been already
published~\cite{kuulkers97}. Here we report preliminary results of the same
observation obtained with MECS and PDS. The on-source exposure times were about
40~ks for MECS and about 20~ks for PDS. Figure~\ref{fig:fig_4}a shows the
source spectrum in the 2--200~keV energy band. The broad-band spectrum is not
well fit by the low-energy model described above (see Fig.~\ref{fig:fig_4}b),
yielding a reduced $\chi^2_\nu$ of 1.62 for 179 degrees of freedom (dof). By adding a
power law component to fit the high energy excess $\chi^2_\nu$
decreases to 1.51. The fit is still not satisfactory and thus a more suitable
model has to be worked out, but this preliminary result shows that a high
energy component is very likely present in the data. The preliminary model
components and fit parameters are the following: soft blackbody with $kT_{bb}
\simeq 1.5$~keV, Sunyaev and Titarchuk~\cite{sunyaev80} Comptonization model
with $kT_{ST} = 3.3$~keV and optical depth $\tau_{ST}\approx 9$~keV, high
energy power law model with photon index $\alpha = 1.9\pm 0.7$.

We estimated a possible contribution from the galactic ridge to the observed
hard tail in the Cygnus X--2 spectrum. Using the results from Yamasaki
et~al.~\cite{yamasaki97}, we conclude that this contribution is a very small
fraction of our 10--100~keV flux.

\subsection{GS1826--238}

GS1826--238 was discovered in September 1988 with {\it Ginga}~\cite{makino88}
at a flux level of 26 mCrab in the 1--40~keV energy band with a hard power law
spectrum (photon index $\Gamma = 1.7$). Later on, the source was detected at a
7$\sigma$ significance level with OSSE above 50 keV with a steep power law
spectrum ($\Gamma = 3.1\pm0.5$)~\cite{strickman96}. The source  was optically
identified with a V19.3 star~\cite{barret95}. Given that its erratic flux time 
variability is reminiscent of that exhibited by Cygnus X--1, the source was 
classified by Tanaka and Lewin~\cite{lewin95} as a black hole candidate. On
March 31, 1997, the Wide Field Cameras (WFC) \cite{WFC} aboard \B\ detected
three X--ray bursts from the source~\cite{ubertini97}, suggesting that we are
in presence of a weak magnetic field neutron star in a LMXRB.

\begin{figure}
\centerline{\epsfig{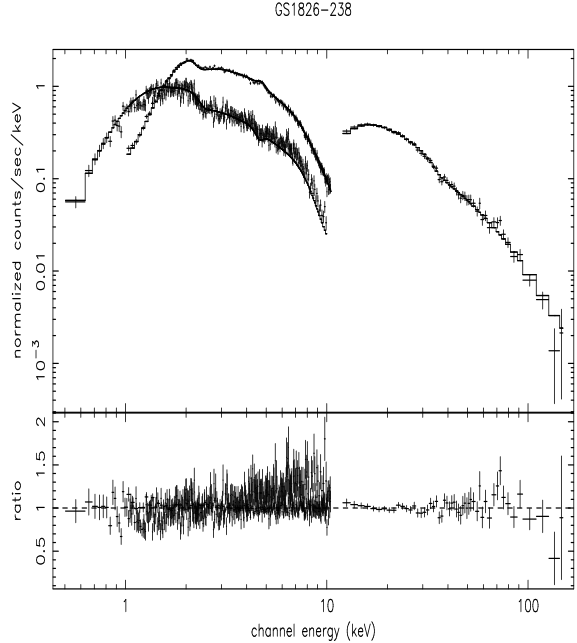}}
\caption[]{Top: Broad band count rate spectrum of GS1826--238 (1st TOO) with
superposed the best fit model (see text). Bottom: residuals with respect to the
model.}
\label{fig:fig_5}
\end{figure}

\B\ again observed the source on October 6--7, 17--18, and 27, 1997 as a target
of opportunity (TOO). The first observation was triggered by a hard X--ray
outburst with a peak flux of about 100~mCrab observed with the BATSE experiment
aboard CGRO. We report here preliminary results obtained during the first
observation. The exposure time was 7.7~ks for LECS, 21.7~ks for MECS and 10~ks
for PDS. The 2--10~keV flux level from the source was $5.8\times
10^{-10}$~\cgs, while at higher energies (20--100~keV) was $7.9\times
10^{-10}$~\cgs. Figure~\ref{fig:fig_5} shows the broad-band count rate spectrum
of the source. A best fit to the data is obtained with an absorbed blackbody
(bb) plus a PL with an exponential cut-off. The best fit parameters are the
following: N$_H = 5.1 \times 10^{21}$~cm$^{-2}$,  $kT_{bb} \approx 1.21$~keV,
PL photon index $\Gamma\approx 1.5$, high energy cut-off parameter
approximately equal to 40~keV. X--ray bursts were observed during both the
first and the second TOOs.

%
% Figure 5
%

\subsection{CYGNUS X--1}

Cygnus X--1 is the most convincing example of a binary system that hides a
stellar mass BH. As discussed above, AGNs are the best candidate objects to
contain massive BH. The Compton reflection  model used for AGNs is suggested to
hold for Cygnus X--1 as well (see, \eg \cite{done92,haardt93}). A test for the
validity of this model for Cygnus X--1 is the presence of a broad Fe K emission
line in the X--ray spectrum of the source as a result of fluorescence from the
disk. This line was actually detected with EXOSAT/GSPC~\cite{barr87} during a
hard X--ray state of the source, but not confirmed in high resolution
observations of Cygnus X--1 with the BBXRT~\cite{marshall93} and
ASCA~\cite{ebisawa96} missions.

%
% Figure 6
%
\begin{figure}
\centerline{\epsfig{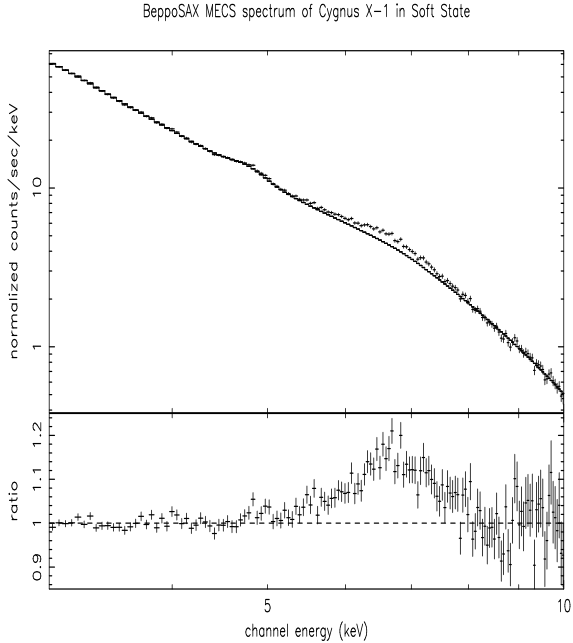}}
\caption[]{Top: Count rate spectrum of Cygnus X--1 measured by MECS during
the June 25, 1996 observations. The other \B\ instruments were
switched off. Superposed is the best fit power law model to the continuum
(excluded from the fit the 4.5--8.5~keV band). Bottom: excesses from
the the model. The broad Fe K line is apparent with no clear evidence of a
K edge.}
\label{fig:fig_6}
\end{figure}                    

\B\ observed the source three times during 1996, on June 22 and 25, during
which the source was in soft state (SS) and on September 12, when the source
was going back to its normal hard X--ray state (HS). Final results of these
observations will be reported elsewhere \cite{frontera97}. Here we report
preliminary results on the Fe K line feature. A broad line is observed in all
the above observations. Figure~\ref{fig:fig_6} shows the count rate spectrum in
the 2--10~keV band measured with the MECS telescopes during the 25 June
observation. Only during the first SS observation a Fe K edge is clearly
detected (E$_{\rm edge} = 7.6$~keV, $\tau_{\rm max} = 0.24$). A reflection
component in the continuum spectrum is detected with the HPGSPC detector
(8.5--60~keV) in both SS and HS. Using a reflection model for the MECS
continuum spectrum around the Fe line, with the reflection parameters estimated
from HPGSPC, one obtains for the spectral feature during the second SS
observation a best fit with a disk-line model with rest energy at 6.4 keV
($\chi^2_\nu = 1.1$). In the other observations, a Gaussian line profile gives
better fits, independently of the continuum used (power law with  or without a
reflection component). The equivalent width is highest in the first observation
(about 1~keV) and lower ($\sim 300$~eV) in the other  observations,
independently of the source state. These results, if interpreted in terms of a
Fe K fluorescence from an accretion disk around a BH, require a different
dimension of the emission region.

\subsection{Mkn 3}

Mkn~3 ($z = 0.0135$) is a type 2 Seyfert first detected in X--rays by
Ginga~\cite{awaki90}. It was only marginally detected above 50~keV with the
OSSE experiment aboard the {\it Compton\/} Gamma Ray
Observatory~\cite{johnson97}.

\begin{figure}
\centerline{\epsfig{figure=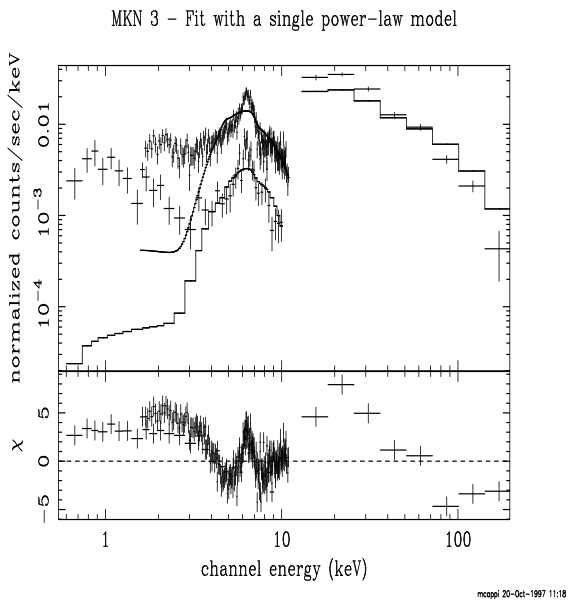,height=8.5cm,width=\columnwidth,angle=-90}}
\caption[]{\B\ spectrum of Mkn~3 fitted with an absorbed power law. The
residuals indicate a strong Fe K line plus a 15--50 keV hump.}
\label{fig:fig_7}
\end{figure}

It was observed with \B\ on November 26, 1996. More details of this 
observation and results can be found elsewhere~\cite{malaguti97}. The flux
level measured in the low (2--10~keV) energy band is $ 6\times 10^{-12}$~\cgs,
corresponding to 0.28 mCrab, while that in the high (20--100~keV) band is $1.3
\times 10^{-10}$~\cgs, corresponding to 7.6~mCrab, with a ratio  (High/Low) =
27. The high energy spectrum has been determined up to 200~keV.
Figure~\ref{fig:fig_7} shows the count rate spectrum of the source in the
0.6-200~keV energy band.  Preliminary results indicate the presence of a soft
excess, a prominent Iron K$\alpha$ complex and a 15--50 keV hump. The heavily
(N$_H \approx 10^{24}$~cm$^{-2}$) absorbed primary ionization provides only a
few percent of the 2--10 keV flux, while it dominates above 10~keV, in the PDS
band. This result can be interpreted in terms of the unified model of 
AGN~\cite{malaguti97}.

%
% Figure 7
%

\section{CONCLUSIONS}

Observations of celestial sources with \B\ show new key results on their high
energy ($>20$~keV) spectral properties.

Strong high energy emission has been detected from Algol during 1 day long
flare. The emission appears to be the tail of the thermal X--ray radiation.

LMXRBs are known sources of low energy X--ray emission, but the high energy
emission is well known only for a small part of them (see a recent review by
\cite{tavani97}). We detected for the first time a non thermal high energy
component from the Z source Cygnus X--2 and derived, for the first time with
the same satellite, the broad--band (0.1--200~keV) photon spectrum of an X--ray
burster during a transient hard X--ray outburst. We expect to be able, using
the three observations available, to study the relative behaviour of the high
energy components with respect to the low-energy one as a function of the high
energy flux of GS1826--238.

We have reported the clear detection of a broad Fe K line feature from Cygnus
X--1 during a soft state of the source and we have discussed its behaviour for
different spectral states. The Fe fluorescence emission does not appear to be
consistent with a constant disk emission region.

As far as the AGNs are concerned, the detection of power laws with strong
absorptions at lower energies and reflection components, is now observed for
several Seyfert 2 galaxies. We have shown the outstanding example of Mkn~3.

\end{document}